\documentclass[reprint, amsmath, amssymb, superscriptaddress, nofootinbib]{revtex4-2}

\usepackage{graphicx}
\usepackage{natbib}
\usepackage{amsmath}
\usepackage{caption}
\usepackage{float}
\usepackage{dcolumn}
\usepackage{bm}
\usepackage[colorlinks=true,citecolor=blue,linkcolor=blue,urlcolor=blue]
{hyperref}
\usepackage{subfig}

\begin{document}
	
\title{On the role of closed timelike curves and \\ confinement structure around Kerr-Newman singularity}


\author{Ayanendu Dutta}
\email{ayanendudutta@gmail.com}
\affiliation{Department of Physics, Jadavpur University, Kolkata-700032, INDIA}

\author{Dhritimalya Roy}
\email{rdhritimalya@gmail.com}
\affiliation{Department of Physics, Jadavpur University, Kolkata-700032, INDIA}

\author{Subenoy Chakraborty}
\email{schakraborty.math@gmail.com}
\affiliation{Department of Mathematics, Jadavpur University, Kolkata-700032, INDIA}


\begin{abstract}
	In this study, the particle motion around the naked singularity and black hole of Kerr-Newman spacetime is investigated with a special attention on the closed timelike orbits. It is found that both in the naked singularity (NS) and in black hole (BH), the singularity is concealed by causality violating regions, and the Cauchy surface consistently resides inside the inner horizon in non-extremal black holes. For neutral particles and particles with an identical charge to the source, only particles with positive angular momentum are permitted to traverse the closed timelike curves. Conversely, for particles with the opposite charge to the source, the strong Coulomb attraction draws all particles inside the Cauchy surface, allowing them to be present in the closed timelike curves irrespective of their angular momentum. However, in both the NS and BH (both extremal and non-extremal), test particles are confined at a considerable distance from the singular point such that there always exists an empty region surrounding the singularity which prevents particles from interacting with it. The radius of the empty surface that depends on the source parameters and the particle characteristics, is investigated with an accurate expression.
\end{abstract}
\maketitle

\section{Introduction}
In Einstein's general relativity, geodesic motion of test particles around the background of black holes is an utterly interesting subject. Astronomical phenomena like black hole shadow, light deflection, perihelion shift of planets, Lense-Thirring, Shapiro effect are related to the geodesic motions. In the coming decades, the LIGO \cite{LIGO}, the Event Horizon Telescope \cite{EHT}, and other high-end experiments \cite{ATHENA,SKA,eLISA} may be able to explore the mysteries of black hole horizon phenomena with a significantly new level of precision.

The Kerr-Newman exact solution of Einstein-Maxwell field equation defines the gravitational field of a charged rotating black hole solution \cite{Newman:1965my}. However, it can also describe a naked singularity for a special limit of the metric parameters. The solution can generalize both the Kerr metric \cite{Kerr:1963ud} and the Reissner-Nordstr\"om metric \cite{reissner1916,nordstrom1918} with specific choices of the angular momentum of the black hole ($ a $) and the electric charge ($ Q $). In the astrophysical context, many of the black hole candidates are expected to have rotation involved with it and thus, the Kerr solution has a particular point of interest. Nevertheless, most of the well-known astrophysical compact objects only possess a small or no net charge, but studies regarding some accretion scenarios may indicate the possibility of black holes with net charge and spin \cite{wilson1975,damour1978,Ruffini:2009hg}. Thus, Kerr-Newman black holes have gained great interest in recent times. The Kerr-Newman solution is important from the phenomenological and conceptual point of view in the sense that it may represent an idealized framework to investigate the interaction between gravitoelectric, gravitomagnetic and electromagnetic components of gravity. Readers are also referred to \cite{Adamo:2014baa} for a recent extended review of the Kerr-Newman metric.

To investigate the gravitational field and the corresponding phenomena in a particular metric, one may explore the geodesic motion of test particles within the spacetime. There are a number of different aspects of geodesic motion studied in Kerr-Newman geometry over time, such as the equatorial timelike and spherical orbits of uncharged test particles \cite{Johnston:1974pn}, last stable orbits of charged particles \cite{Young:1976zz}, general discussions of radial motion, motion along the symmetry axis, motion of ultrarelativistic particles and zero angular momentum particles \cite{bicak1989,balek1989,stuchlik1999}, unstable circular orbits of charged particles outside the outer horizons \cite{Kovar:2008ta} and so on \cite{Sharp:1979sqa}. Alongside, Calvani and Turolla comprehensively investigated the photon orbits and the extended manifold with negative radial coordinate and naked singularity \cite{Calvani:1981ya}. Some recent studies has also been focused on the analytic solution of Kerr-Newman geodesics in terms of Weierstrass elliptic functions to differentiate various orbits of charged particle motion \cite{Hackmann:2013pva}, the geodesics in (A)dS and in $ f(R) $ modified gravity model \cite{Soroushfar:2016esy}. Pugliese \textit{et al.} \cite{Pugliese:2013zma} studied the equatorial orbits of neutral test particles with a special attention to distinguish between black hole and naked singularity. They adopted the effective potential approach to examine various regions of orbits in both the black hole and naked singularity. The important point however coming out of this discussion is the empty region surrounding the central singularity which prevents test particles from interacting with the singular point. This region has some significant relevance with the present study. For other recent investigations that particularly focus on the particle motion and geodesics of extremal black hole, or related aspects of Kerr-Newman geometry, readers are referred to \cite{Stuchlik:2000pla,Liu:2017fjx,Wang:2022ouq,Hsiao:2019ohy,Slany:2020jhs,Kraniotis:2019ked,Galtsov:2019bty,Garnier:2023lph,Kapec:2019hro,Wei:2016avv,Chen:2016caa,Xu:2014seo,Chrusciel:2019xuf}.

Einstein's GR, on the other hand, comes with an interesting paradox around the singularity of spacetime geometry, called the closed timelike curves (CTCs) that allow `time' to flow in the backward direction \cite{Hawking:1991nk}. On these orbits, the order of time, i.e. chronology, is violated, leading to the loss of determinism. The nature of the paradox involved in the CTC is different from that of the curvature singularities, as the singularity causes the breakdown of laws of physics where the CTC indicates the breakdown of predictability. However, likewise the singularity problem, the spacetime region that causally connects the CTCs (i.e. the causality violating region), leads to the ill defined Cauchy problem of initial value. The surface of this region is therefore commonly known as the `Cauchy horizon' \cite{Visser:1995cc}. Nevertheless, there are attempts to successfully overcome the appearance of CTC by introducing modified versions of certain spacetime, for example \cite{Prasad:2018hdj}, it still isn't completely described \cite{Bojowald:2005qw,Mbonye:2005im,Hossenfelder:2009fc,Martin-Vazquez:2018usf}.

The theory behind the physics of time travel with CTC itself contains pathology such as inconsistency paradox and grandfather paradox etc. Nevertheless, recent study by Tobar and Costa has proposed that paradox-free (or consistency-free) time travel is possible around CTCs \cite{Tobar:2020ybp}. They have found that in non-trivial time travel, multiple inequivalent processes may exist together which supports the complex dynamics around CTCs leading to the consistency-free operations. Recently, Nolan studied the motion of a gyroscope on closed timelike curves and investigated $ T $-periodic spin-vectors. For different spacetimes admitting CTCs, they discussed the results from the consistency-principle perspectives \cite{Nolan:2021ywy}.

The CTCs occur from the causality violating trajectories which is a generic feature of most of the stationary axially symmetric rotating solutions of General Relativity, first being the G\"odel's cosmological solution \cite{Godel:1949ga} that was first presented in 1949. Later, it is found that other spacetimes like Kerr metric \cite{Kerr:1963ud}, van Stockum rotating dust \cite{vanStockum:1937zz}, Tipler's cylinder \cite{Tipler:1974gt}, Bonnor's rotating dust cloud \cite{bonnor1977}, Gott's cosmic string \cite{Gott:1990zr} also admits the presence of such curves. For recent works on CTCs, readers are advised to go through \cite{Rovelli:2019ltw,Bishop:2020qtt,Ahmed:2020idw,Luminet:2021qae,Hazarika:2021iuo,Ahmed:2021dkw,Duan:2021pci,Gutti:2021fdo,Vairogs:2021zhv,Barcelo:2022jbr,Krasnikov:2022koq,Gutti:2022qsv,Baladron:2023oya,Zhao:2023tox,Sanzeni:2023qxm,Astesiano:2023emg,Nguyen:2023fzk,Dutta:2024luw}. The Kerr-Newman spacetime, although a modified version of Kerr metric, is an important candidate in this context which originates from the Einstein-Maxwell action. There are a very few studies that certainly investigates the CTC in Kerr-Newman metric \cite{Chrusciel:2012gz}, though, we must mention the seminal works of Brandon Carter on Kerr metric, known as Carter's time mechine \cite{Carter:1966zza,Carter1968}. Due to the possible presence of gravitoelectric and gravitomagnetic components, the CTC in the Kerr-Newman geometry stood out from others, and therefore, contains special theoretical significance. At the same time, the interrelation between CTC and those components of gravity seems to be worthy of more extended discussions. On the other hand, to discuss the complete structure of the spacetime geometry in a more significant way, the characterization of the pathology (i.e. CTC) is indeed necessary to investigate. Hence, motivated by the above discussions, the characteristics of particles around the CTC is intended to explore in terms of geodesic motion of different test particles.

The method used in the study is based on the spacetime diagrams that are largely used in the incoming and outgoing null geodesics to analyze the geometry of a black hole. Among numerous literatures, we refer to a recent study by Cruz \textit{et al.} \cite{Cruz:2013ufa} where the geodesic structure of 2+1 dimensional Lifshitz black hole is investigated mostly on the basis of spacetime diagrams. Corresponding to the CTC in the axially symmetric, stationary, rotating spacetime geometries, the spacetime diagram and respective geodesic confinements are investigated in van Stockum spacetime in \cite{Dutta:2022xew}. Here, we adopt the same approach to examine the geodesic motion, their corresponding confinements and the characteristics of particles inside the CTC in terms of different angular momentum of test particles ($ L $) in Kerr-Newman spacetime.

Hence, in the study, the sections are arranged as follows: in section \ref{geometry}, we summarize the Kerr-Newman geometry in brief, particularly focusing on the horizon structures and the position of CTC. We calculate the geodesic equations of neutral particles in section \ref{geodesics}, where section \ref{naked-sing} and \ref{black-hole} are respectively dedicated for the detailed analysis of geodesic motion and the characteristics of neutral test particles within CTC in naked singularity and black hole (both non-extremal and extremal black holes). Next, in section \ref{charged}, we compile the ideas and put them in motion of charged particles. Finally, we end the study by presenting the concluding remarks in section \ref{discussions}. 

\section{Kerr-Newman Geometry}\label{geometry}
The asymptotically flat, axially symmetric, stationary Kerr-Newman spacetime can be effectively found out from the solution of Einstein-Maxwell field equation given by
\begin{equation}
	G_{\mu \nu}= -2 \left( g^{\alpha \beta} F_{\mu \alpha} F_{\nu \beta} -\frac14 g_{\mu \nu} F_{\alpha \beta} F^{\alpha \beta} \right),
	\label{2.1}
\end{equation}  
where $ G_{\mu \nu} $ and $ F^{\mu \nu} $ is the Einstein tensor and electromagnetic tensor respectively. The Boyer-Lindquist form of the metric is written as
\begin{eqnarray}
	\nonumber	ds^2 = \frac{\rho^2}{\Delta}dr^2 + \frac{sin^2 \theta}{\rho^2}\left[(r^2+a^2)d\phi -a dt^2 \right]^2 \\ -\frac{\Delta}{\rho^2}\left[a sin^2 \theta d\phi -dt \right]^2 +\rho^2 d\theta^2,
	\label{2.2}
\end{eqnarray}
where
\begin{eqnarray}
	\Delta &=& r^2-2Mr+a^2+Q^2,
	\label{2.3}\\
	\rho^2 &=& r^2+a^2 cos^2\theta.
	\label{2.4}
\end{eqnarray}

Here, $ M $ is the mass (always taken to be greater than zero), $ a $ is the angular momentum of the spacetime, and $ Q $ is the electric charge of the gravitational source. The description of the metric reduces to the Schwarzchild metric when $ Q=a=0 $, and at the same time, the Kerr and Reissner-Nordstr\"om metrics are recovered by considering $ Q=0 $ and $ a=0 $ respectively.

The horizons exist in the spacetime when we consider $ \Delta=0 $, and get $ r_{\pm}=M \pm \sqrt{M^2-a^2-Q^2} $. Here, $ r_+ $ and $ r_- $ are respectively the outer horizon (which is the event horizon for Kerr-Newman black hole) and inner horizon. However, the null coordinate expression of the metric does not involve singularity at $ \Delta=0 $. Although the physical significance of the coordinate singularity is important in the context of the geometry, the only genuine singularity of the spacetime is present at $ r=0 $.

\begin{figure}[!]
	\centerline{\includegraphics[scale=.45]{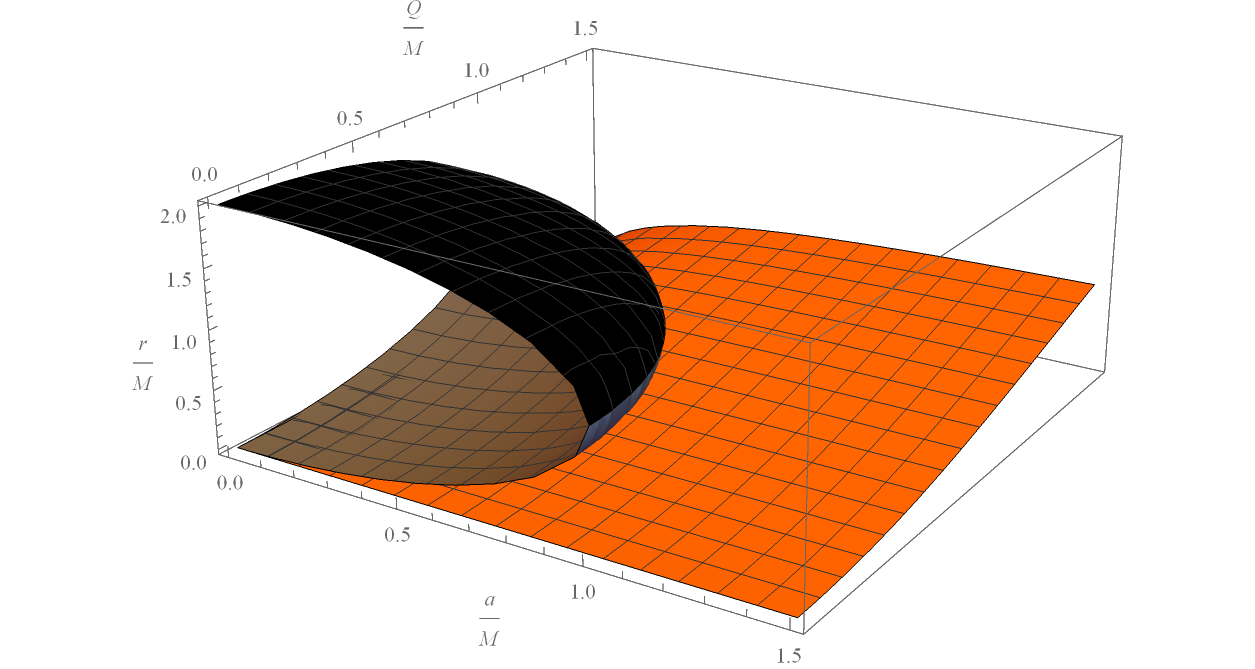}}
	\caption{The boundary of CTC $ r^-_+ $ (orange surface), the inner (gray surface) and outer (black surface) horizons are plotted as a function of $ a/M \in (0,1.5) $ and $ Q/M \in (0,1.5) $. For Kerr-Newman black holes, $ r^-_+ $ always lies inside the event horizon, and in Kerr-Newman naked singularity the existence of CTC is open and it covers the central singularity at $ r=0 $.}
	\label{ctc-fig}
\end{figure}

At this point, the following three observations can be drawn from the expression of $ r_{\pm} $ \cite{Adamo:2014baa}

\begin{enumerate}
	\item For $ M^2>(a^2+Q^2) $, both the inner and outer horizons exist, and the singularity is hidden behind the event horizon. The interior of the black hole given by $ r<r_+ $ is hidden to an observer at infinity.
	\item For $ M^2=(a^2+Q^2) $, the inner and outer horizons coincide at $ r=M $. It describes the so-called Kerr-Newman extremal black hole.
	\item For $ M^2<(a^2+Q^2) $, the event horizon is absent and the geometry is a naked singularity. The important aspect of this particular choice is the violation of causality, where the closed timelike orbit is not hidden inside an event horizon.
\end{enumerate}

However, to describe a closed timelike curve in a stationary, axisymmetric spacetime, one may consider $ g_{\phi \phi} $ to be less than zero along with constant $ r, \theta, t $ \cite{Visser:1995cc, Lobo:2008leb}. For Kerr-Newman metric, eventually it takes the form
\begin{equation}
	r^4+a^2(r^2+2Mr-Q^2)<0,
	\label{2.5}
\end{equation}
for a curve in the equatorial slice of the geometry. These regions have boundaries which can be explicitly obtained from the solution of $ r^4+a^2(r^2+2Mr-Q^2)=0 $. The radii given by the roots are
\begin{eqnarray}
	r^{+}_{\pm} &\equiv& \frac12 \sqrt{\zeta} \pm \frac12 \sqrt{-2a^2-\zeta-\frac{4 a^2 M}{\sqrt{\zeta}}}, \label{ctc1} \\
	r^{-}_{\pm} &\equiv& - \frac12 \sqrt{\zeta} \pm \frac12 \sqrt{-2a^2-\zeta+\frac{4 a^2 M}{\sqrt{\zeta}}}, \label{ctc2}
\end{eqnarray}
where
\begin{widetext}
	\begin{eqnarray}
		\nonumber	\zeta \equiv -\frac{2a^2}{3} + \frac{2^{1/3} (a^4-12a^2 Q^2)}{3 \Big(2a^6+108a^4M^2+72a^4Q^2+ \sqrt{-4(a^4-12a^2Q^2)^3 +(2a^6+108a^4M^2+72a^4Q^2)^2} \Big)^{1/3}} \\
		+\frac{\Big(2a^6+108a^4M^2+72a^4Q^2+ \sqrt{-4(a^4-12a^2Q^2)^3 +(2a^6+108a^4M^2+72a^4Q^2)^2} \Big)^{1/3}}{3 \times 2^{1/3}}.
		\label{ctc3}
	\end{eqnarray}
\end{widetext}

So, for a CTC in Kerr geometry, only Eq. \eqref{ctc3} changes value with $ Q=0 $, leaving behind Eq. \eqref{ctc1} and \eqref{ctc2} invariant. Further, we see that for a given value of source parameters (i.e., $ a,~Q,~M $), $ \zeta $ is always positive, so as $ M $. Thus, $ r^+_\pm $ always produce complex radii. On the other hand, $ r^-_+ $ and $ r^-_- $ respectively gives real positive and real negative values. The behaviour of the $ r^-_+ $ along with the inner and outer horizons are illustrated in Fig. \ref{ctc-fig}. One can notice that up to the point where inner and outer horizons coincide, $ r^-_+ $ (boundary of CTC) always lies inside the horizons. Beyond that, the value of source parameters $ a/M $ and $ Q/M $ define naked singularity, and the horizons are absent. Alongside, as $ Q/M $ decreases, the CTC region moves more towards the central singularity, and for $ Q/M \rightarrow 0 $, it tends to coincide the singularity at $ r=0 $. The same phenomenon occurs also for $ a/M \rightarrow 0 $. So, individually for Kerr and Reissner-Nordstr\"om, the causality violating region lies very much close to the singularity. However, the Cauchy surface denoted by the orange surface always lies inside the inner horizon (gray surface), representing the kind of motion a particle can go through inside the inner horizons of rotating black hole.

\section{The Equatorial Geodesics}\label{geodesics}
To work out the geodesic equations of the metric \eqref{2.2}, one may use the Lagrangian formulation. We restrict ourselves to the equatorial geodesic motion in the present study, hence $ \theta $ is taken to be $ \pi/2 $, and for the obvious reason, the geodesic equation corresponding to $ \theta $ is absent throughout the study. The canonical momentum corresponding to the Euler-Lagrange dynamics provides us the first order geodesic equations given by
\begin{widetext}
	\begin{eqnarray}
		\dot{t} &=& \frac{E (r^4+a^2(r^2+2Mr-Q^2))-aL(2Mr-Q^2)}{r^2(r^2+a^2-2Mr+Q^2)},
		\label{2.6}\\
		\dot{\phi} &=& \frac{aE (2Mr-Q^2)+L(r^2-2Mr+Q^2)}{r^2(r^2+a^2-2Mr+Q^2)},
		\label{2.7}\\
		\dot{r}^2 &=& \frac{(E^2-\mu)r^4+2Mr^3\mu-(-a^2E^2+L^2+a^2\mu+Q^2\mu)r^2+(-aE+L)^2(2Mr-Q^2)}{r^4}.
		\label{2.8}
	\end{eqnarray}
\end{widetext}

Here, the constants of motion $ E=-\frac{\partial \mathcal{L}}{\partial \dot{t}}=-g_{\alpha \beta} \xi^\alpha_t u^\beta $ and $ L=\frac{\partial \mathcal{L}}{\partial \dot{\phi}}=g_{\alpha \beta} \xi^\alpha_\phi u^\beta $ are respectively associated with the total energy and angular momentum of the test particle with mass $ \mu $. The killing vectors $ \xi_t=\partial t $ and $ \xi_\phi=\partial \phi $ are timelike and spacelike in nature and respectively represent the stationarity of the space and the axial symmetry of the source.

A number of observations can be drawn at this point to discuss the co-rotation effect and circular motion of test particles in spacetime. From the $ \phi $ geodesic equation, if we impose $ L=0 $ in Eq. \eqref{2.7}, the angular velocity $ \dot{\phi} $ of the test particles are still not zero, having expression
\begin{equation}
	\dot{\phi}= \frac{aE(2Mr-Q^2)}{r^2(r^2+a^2-2Mr+Q^2)}.
	\label{2.17}
\end{equation}

This can be interpreted as the spacetime dragging. So, while we consider $ L=0 $, the angular momentum-less particles still co-rotates with the spacetime, and if the angular momentum of the spacetime vanishes i.e. $ a=0 $, the dragging effect vanishes.

The possibility of circular motion of test particles can be represented by adopting the classical effective potential approach. The test particle dynamics can be thought of as the one-dimensional classical particle motion in effective potential $ V(r) $. The effective potential in equatorial timelike geodesic can be written as \cite{Misner:1973prb}
\begin{equation}
	V=\frac{-B \pm \sqrt{B^2-4AC}}{2A},
	\label{2.13}
\end{equation}
where $ A,~B $ and $ C $ are directly obtained from the radial timelike geodesic equation (i.e. Eq. \eqref{2.8}), and are given by 
\begin{eqnarray}
	A &=& r^4+a^2(r^2+2Mr-Q^2),
	\label{2.14}\\
	B &=& -2aL(2Mr-Q^2),
	\label{2.15}\\
	\nonumber
	C &=& -\mu r^4+2Mr^3\mu-(L^2+a^2\mu+Q^2\mu)r^2\\
	&&+L^2(2Mr-Q^2).
	\label{2.16}
\end{eqnarray}

We can obtain the circular orbits of test particles by the simultaneous solutions of the equations
\begin{eqnarray}
	V'(r,L,a,Q)=0, ~~~ \text{and} ~~~ V=E/\mu.
	\label{2.16a}
\end{eqnarray}
where $ ' $ denotes differentiation with respect to $ r $.

The main aim of this study is to characterize the motion of test particles corresponding to the black hole and naked singularity as a function of angular momentum ($ L $). So, in order to explore the dynamics of circular motion, we analyze the condition $ V'=0 $, and solve it with respect to the angular momentum (AM). The general solution is given by
\begin{eqnarray}
	\frac{L_{\pm}}{\mu }\equiv \frac{1}{r^2}\sqrt{\frac{\mathcal{X} \pm2M^2 \sqrt{\mathcal{Y}}}{\mathcal{Z} }},
	\label{2.16b}
\end{eqnarray}
where $ L=\pm L_{\pm} $, and
\begin{widetext}
	\begin{eqnarray}
		\nonumber	\mathcal{X}&\equiv& r^2 \left\{-\left(Q^2-Mr\right) r^4 \left[2 Q^2+(r-3M) r\right]+a^4 \left(Q^2-Mr\right) \left[2 Q^2-(5M+r) r \right]+\right.
		\\
		&&\left. a^2 \left[2 Q^6+Q^4 (r-11M) r-2 Q^2 (r-2M) r^2 (5M+r)+2M r^3 [r (3M+r)-6M^2]\right]\right\},
		\label{2.16c}	
		\\
		\mathcal{Y}&\equiv&-a^2 \left(Q^2-rM\right) r^4 \left[a^2+Q^2+(r-2M) r\right]^2 \left[a^2 \left(Q^2-Mr\right)+\left(2 Q^2-3Mr\right) r^2\right]^2,
		\label{2.16d}
		\\
		\mathcal{Z}&\equiv&4 a^2 \left(Q^2-rM\right)+\left[2 Q^2+(r-3M) r\right]^2 .
		\label{2.16e}
	\end{eqnarray}
	
	The corresponding energy is obtained from Eq. \eqref{2.16a}, which takes the form
	\begin{eqnarray}
		E({\pm L_{\mp}})=\frac{- (\pm L_{\mp}) \left(Q^2-2 M r\right)+\sqrt{r^2 \left[a^2+Q^2+(r-2M) r\right] \left(r^2 \left(L_{\mp}^2+r^2\right)+a^2 \left[r (2M+r)-Q^2\right]\right)}}{r^4+a^2 \left[r (2M+r)-Q^2\right]}.
		\label{2.16f}
	\end{eqnarray}
\end{widetext}

Here, $ L_+ ~\text{and}~ L_- $ respectively denotes the positive and negative segment of the angular momentum. The well defined region of energy and angular momentum in the orbital coordinate and source parameters are briefly investigated in \cite{Pugliese:2013zma} by Pugliese \textit{et al.} They also analyzed the orbital regions where circular motion occurs. However, the analysis of the region of space of circular orbits is not particularly necessary in the present study. But, the interrelation between the effective potential and the angular momentum is important.

The main aim of this study is to find the region of space where geodesic trajectories and closed timelike curves within a Cauchy horizon is effectively possible. To establish the motion around CTC in Kerr-Newman geometry, we adopt a number of simple techniques such as geodesics in spacetime diagrams, the velocity analysis and effective potentials in radial timelike geodesics. It is evident from textbook discussions that CTC appears in some well-known rotating space-times at a certain region of space. On the other hand, regardless of complete geodesic structure in a metric, there are certain confinements possible for particles with different characteristics. For the spacetime diagram i.e. $ r-t $ plot, the expression for $ dt/dr $ is introduced from the geodesic equations \eqref{2.6}, and \eqref{2.7}. The trajectories are discussed corresponding to zero, positive and negative values of AM of test particles. The particles with positive AM are such particles which co-rotate with the spacetime (i.e. the direction of rotation is same as the spacetime), negative AM particles counter-rotate with the spacetime, and zero AM particles only rotate with the spacetime dragging. On the other hand, without loss of generality, we will restrict ourselves to the case of positive charge and positive angular momentum of the spacetime (i.e. $ Q>0 $ and $ a>0 $). We start the discussion with the horizon-less naked singularity followed by black holes with inner and outer horizons.

\section{Naked Singularity}\label{naked-sing}

In this section, we shall discuss the motion of neutral test particles around the KN naked singularity in detail. Focusing on the effective potential and the motion of angular momentum-less particles, the solution of the equations
\begin{eqnarray}
	V'(r,L,a,Q)=0, ~~~ \text{and} ~~~ L=0,
	\label{naked1}
\end{eqnarray}
marks the particular circular motion in the geometry regarding the real solution that only exists in the naked singularity. In general, the gravitational component of the effective potential governs the motion represented by Eq. \eqref{naked1}, and $ L=0 $ motion is possible only when there is a balance of forces in the configuration. So, referring to \cite{felice1974}, it can be interpreted by the repulsive gravity effect.

Further, the characterization of particle motion with non-zero angular momentum is more complex in KN naked singularity. So, in order to establish the necessary arrangements required for the study, the corresponding radial null geodesic equations in terms of $ dt/dr $ for naked singularity are written as
\begin{widetext}
	$\bullet$ \textbf{Geodesics with zero angular momentum:}
	\begin{eqnarray}
		\frac{dt}{dr} =\pm \frac{\sqrt{r^4+a^2(r^2+2Mr-Q^2)}}{(r^2+a^2-2Mr+Q^2)},
		\label{2.10}
	\end{eqnarray}
	
	$\bullet$ \textbf{Geodesics with non-zero angular momentum:}
	\begin{eqnarray}
		\frac{dt}{dr} =\pm \frac{E (r^4+a^2(r^2+2Mr-Q^2))-aL(2Mr-Q^2)}{(r^2+a^2-2Mr+Q^2) \sqrt{E^2r^4-(-a^2E^2+L^2)r^2+(-aE+L)^2(2Mr-Q^2)}}.
		\label{2.9}
	\end{eqnarray}
\end{widetext}

\begin{figure*}[t]
	\centering
	\subfloat[\label{NSNG-fig}]{{\includegraphics[scale=0.55]{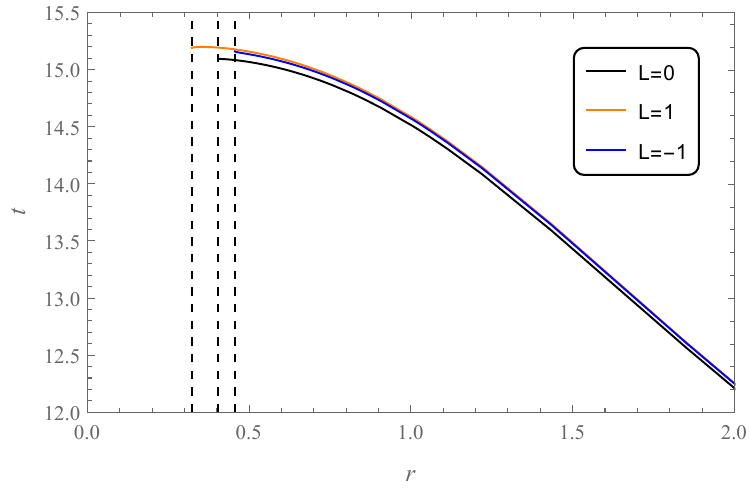}}}\qquad
	\subfloat[\label{NSTG-fig}]{{\includegraphics[scale=0.55]{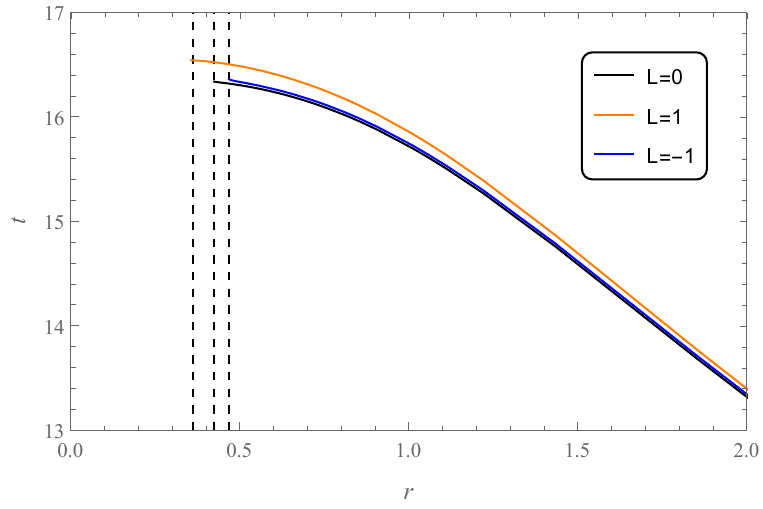}}}
	\caption{(a) Spacetime diagram for radial null geodesics of naked singularity. The trajectories for zero and non-zero AM photons are plotted with $ E=2 $, and $ a=Q=M=1 $. $ L=0,~L=1, $ and $ L=-1 $ trajectories are respectively confined at $ r=0.405,~r=0.322, $ and $ r=0.456 $, which are shown with black dashed lines. (b) Spacetime diagram for radial timelike geodesics of naked singularity. The trajectories for zero and non-zero AM massive particles are plotted with $ E=2,~\mu=1 $, and $ a=Q=M=1 $. $ L=0,~L=1, $ and $ L=-1 $ trajectories are respectively confined at $ r=0.424,~r=0.362, $ and $ r=0.468 $, which are shown with black dashed lines.}
	\label{NS_spacetime-fig}
\end{figure*}

The spacetime diagram associated with the numerical integration of Eq. \eqref{2.10} and \eqref{2.9} are exhibited in Fig. \ref{NSNG-fig}. For positive and negative values of angular momentum, $ L=1 $ and $ L=-1 $ is chosen along with the photon energy $ E=2 $, and naked singularity source parameters $ a=Q=M=1 $. For the same parameter choices, closed timelike curves appear at the orbital radius $ r<0.405 $ which can be directly obtained from $ r^-_+ $ in Eq. \eqref{ctc2}.

From Eq. \eqref{2.10}, it is convenient to note that the trajectories of angular momentum-less photons do not depend on the associated particle energy.
It is readily visible from the figure that the particles are forbidden to interact with the central singularity at $ r=0 $. Their trajectories are always confined at a particular position significantly far from the singular point.

Further, referring to Fig. \ref{NSNG-fig}, notice that, the zero and negative ($ L=-1 $) AM photons are confined respectively at $ r=0.405, $ and $ r=0.456 $, and they are forbidden to approach $ r<0.405 $ position where CTC appears. Thus, one may comfortably conclude that zero and negative AM photons cannot traverse the CTCs. However, positive AM ($ L=1 $) photons can pass through $ r=0.405 $ to reach $ r=0.322 $, indicating their presence in CTC.

\subsection{Timelike motion}
For timelike massive ($ \mu \ne 0 $) particles, the associated geodesic equations are written as

\begin{widetext}
	$\bullet$ \textbf{Geodesics with Zero angular momentum:}
	\begin{eqnarray}
		\frac{dt}{dr} =\pm \frac{E (r^4+a^2(r^2+2Mr-Q^2))}{(r^2+a^2-2Mr+Q^2) \sqrt{(E^2-\mu)r^4+2Mr^3\mu-r^2\mu(a^2+Q^2)+a^2E^2(r^2+2Mr-Q^2)}}.
		\label{2.12}
	\end{eqnarray}
	
	$\bullet$ \textbf{Geodesics with Non-zero angular momentum:}
	\begin{eqnarray}
		\frac{dt}{dr} =\pm \frac{E (r^4+a^2(r^2+2Mr-Q^2))-aL(2Mr-Q^2)}{(r^2+a^2-2Mr+Q^2) \sqrt{(E^2-\mu)r^4+2Mr^3\mu-(-a^2E^2+L^2+a^2\mu+Q^2\mu)r^2+(-aE+L)^2(2Mr-Q^2)}}.
		\label{2.11}
	\end{eqnarray}
\end{widetext}

\begin{figure}[t]
	\centering
	\subfloat[\label{NSEP-fig}]{{\includegraphics[scale=0.55]{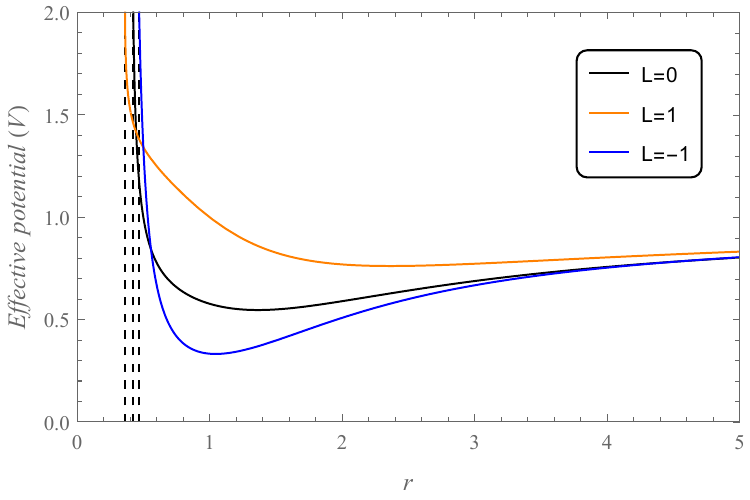}}}\qquad
	\subfloat[\label{NS_time-fig}]{{\includegraphics[scale=0.55]{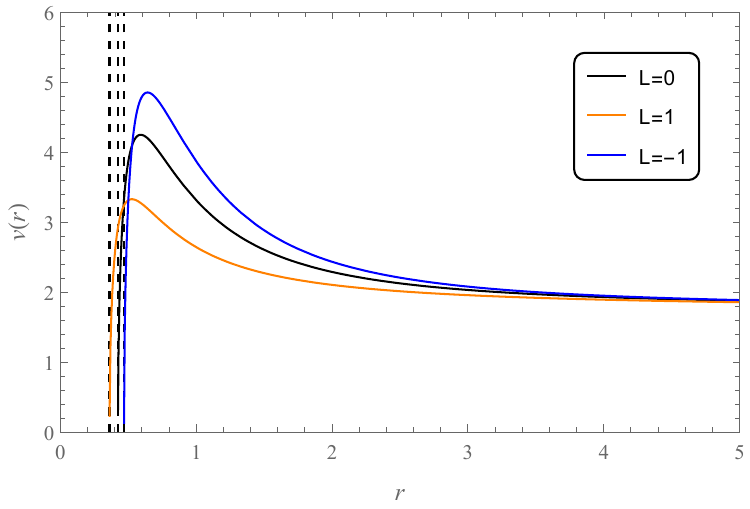}}}
	\caption{(a) Effective potential and the (b) velocity in terms of proper time of zero and non-zero AM test particles are exhibited with $ E=2,~\mu=1 $, and $ a=Q=M=1 $. The points where the plots drop to positive infinity or to zero are shown in black vertical dashed lines which accurately coincide with the confinement points of Fig. \ref{NSTG-fig}.}
\end{figure}

The spacetime diagram of Eq. \eqref{2.12} and \eqref{2.11} for massive particles with zero and non-zero AM are plotted for the same choices of parameters, i.e. $ a=Q=M=1 $ and $ E=2 $, in Fig. \ref{NSTG-fig}. For non-zero angular momentum, $ L=1 $ and $ L=-1 $ is chosen. These trajectories also include confinement points shown by vertical black dashed lines which can be obtained with the same approach as the previous case. However, these particles penetrate spacetime less than the photons due to their inbound masses. Additionally, from the plot, one can distinguish the particle movement inside a closed timelike loop. Only the test particles with positive AM can pass through the $ r=0.405 $ boundary, thus making their way into the CTC.

Now, to represent the confinement phenomenon in a physical context, one can adopt the effective potential approach. A plot of the effective potential from Eq. \eqref{2.13} is shown in Fig. \ref{NSEP-fig} for the same parameter choices as Fig. \ref{NSTG-fig}. It is observed from the figure that the singularity is covered by the classical effective potential which asymptotes to positive infinity at the confinement point such that it restricts timelike geodesics to reach the singularity.

In the context of the gravitational field, we would expect that an infalling particle in the naked singularity which leaves from spatial infinity would enter regions of increasingly higher gravitational field as it moves towards the singularity at the symmetry axis. Gradually it would feel the maximum effect at the minima of effective potential, and would then continue towards decreasing gravitational field. Explaining it another way, the infalling particles reach the maximum gravitational attraction at the minima of Fig. \ref{NSEP-fig}, and then are repelled by the field at the confinement point or the inner turning point. The phenomenon seems to be simple, and is in fact represented by the nature of light cones on the axial geometry. Qualitatively, the inner region up to the point of minima acts itself as the source of gravity, and the spacetime curvature on the axial symmetry drops to zero as $ r \rightarrow \infty $.

For a complete task on the discussion, the velocity in terms of proper time attained by the infalling particle is exhibited as a function of radial parameter in Fig. \ref{NS_time-fig} by using the radial four-velocity expression given by Eq. \eqref{2.8}. It is rather straightforward to observe that the particles attain maximum velocity at the minimum of the effective potential where they would feel the maximum field. Then as they move more towards the singularity, the peculiar nature of the geometry repels down the particles up to zero velocity. Finally, it is evident that the lower minima of the potential attracts and repels the positive AM particles more loosely than the other two, so that they penetrate the spacetime more to reach the CTC region leaving behind the zero and negative AM particles. 

\section{Black Hole}\label{black-hole}

\begin{figure*}[t]
	\centering
	\subfloat[\label{BHNG-fig1}]{{\includegraphics[scale=0.39]{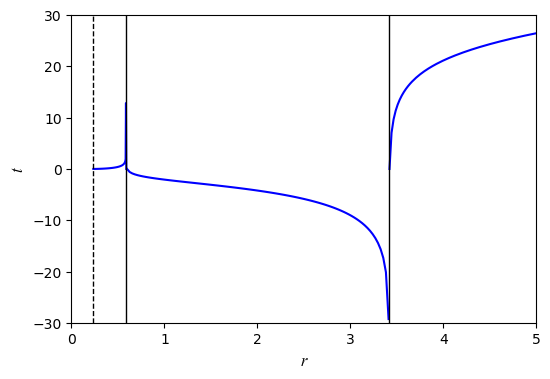}}}\qquad
	\subfloat[\label{BHNG-fig2}]{{\includegraphics[scale=0.39]{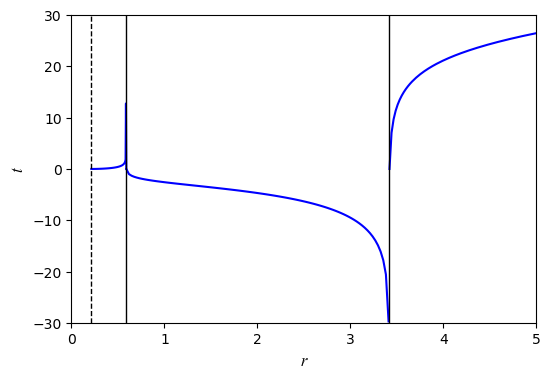}}}\qquad
	\subfloat[\label{BHNG-fig3}]{{\includegraphics[scale=0.39]{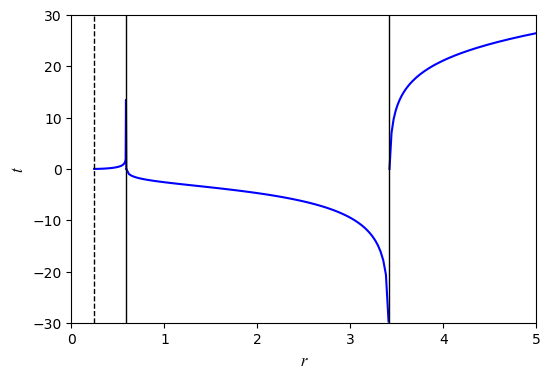}}}\\
	\subfloat[\label{BHTG-fig1}]{{\includegraphics[scale=0.39]{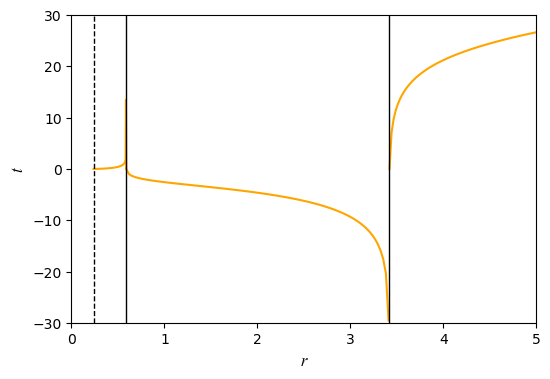}}}\qquad
	\subfloat[\label{BHTG-fig2}]{{\includegraphics[scale=0.39]{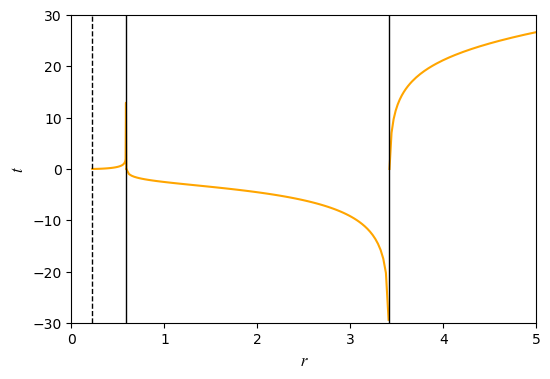}}}\qquad
	\subfloat[\label{BHTG-fig3}]{{\includegraphics[scale=0.39]{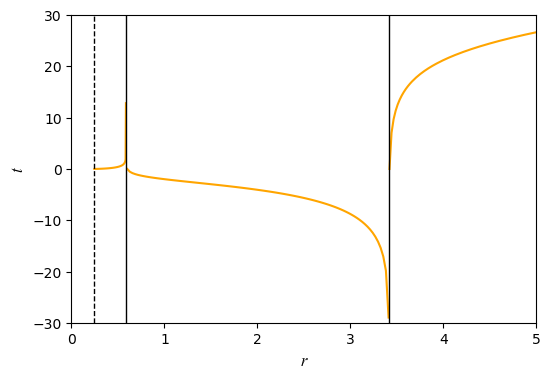}}}
	\caption{TOP PANEL: Spacetime diagram for radial null geodesics of KN black hole with $ r_{ctc}=0.235,~r_{in}=0.586, $ and $ r_{out}=3.414 $. The plots are exhibited for $ E=2,~a=Q=1,~ M=2 $, and are separated with respect to angular momentum of photons as (a) $ L=0 $, (b) $ L=1 $, and (c) $ L=-1 $. The confinement points are shown in black dashed lines, and are located at $ r=0.235,~r=0.214,~ \text{and}~ r=0.245 $ respectively. \newline BOTTOM PANEL: Spacetime diagram for radial timelike geodesics of KN black hole with $ r_{ctc}=0.235,~r_{in}=0.586, $ and $ r_{out}=3.414 $. The plots are exhibited for $ E=2,~\mu=1,~a=Q=1,~ M=2 $, and are separated with respect to total angular momentum as (d) $ L=0 $, (e) $ L=1 $, and (f) $ L=-1 $. The confinement points are shown in black dashed lines, and are located at $ r=0.238,~r=0.224,~ \text{and}~ r=0.246 $ respectively. In each plot, the left and right black vertical solid lines respectively denote the inner, and outer horizons.}
	\label{BH-fig}
\end{figure*}

In this section, we shall discuss the existence of CTC and the nature of particles traversing them in Kerr-Newman black holes with inner and outer horizons. CTC present in the Kerr-Newman black hole appears inside the inner horizon, thus allowing its presence undetectable for the observers at infinity. At the same time, making its way into the CTC for a particle is relatively difficult in the sense that it must be crossing two horizons where spaghettification may be obvious. Still, the study of geodesic behavior and CTC in rotating black holes is important as it may extend the knowledge of inner structure.

Notice that, the basic geodesic equations for the discussion of radial geodesics in KN black holes are the same as the naked singularity, however, for the choices of source parameters, $ M^2>(a^2+Q^2) $ condition must be satisfied.

Regardless of the presence of inner and outer horizons in KN black hole, the spacetime diagram for the radial null geodesics with respect to an observer at infinity, is shown in the top panel of Fig. \ref{BH-fig} (by numerically integrating Eq. \eqref{2.10} and \eqref{2.9}), where the motion of $ L=0,~L=1 $ and $ L=-1 $ are exhibited separately for $ E=2,~ a=Q=1, $ and $ M=2 $. For these source parameter choices, the CTC, inner, and outer horizons appear in the orbital radius at $ r_{ctc}=0.235,~r_{in}=0.586, $ and $ r_{out}=3.414 $ respectively. 

It is obvious that the motion of particles in KN black hole is not straightforward like the naked singularity. The trajectories appear to consume infinite times at the horizons, as seen from the figures. It is well-known across textbook literature that in Schwarzschild black hole, the geodesics of infalling particles consume infinite times to intercept the event horizon, however, with respect to proper time the particles hit the singularity with finite time \cite{Hawking:1973uf,Misner:1973prb,dInverno:1992gxs}. An exactly same phenomenon is observed in the inner and outer horizons of Kerr-Newman black hole, only that the trajectories are exempted from approaching the central singularity at the confinement points. Nevertheless, one fact is sure, similar to the naked singularity, all the three motions are confined at particular points inside the inner horizon, although they are significantly far from the singular point at $ r=0 $. Additionally, the characteristics of particles in CTC are similar to the naked singularity. The $ L=0 $ motion is confined at the limiting radius of CTC, where the negative AM photons do not approach the radial boundary where CTC exists. Thus, zero and negative AM particles are forbidden in closed timelike orbits.

\subsection{Timelike motion}

\begin{figure}[t]
	\centering
	\subfloat[\label{BHTG-proper}]{{\includegraphics[scale=0.55]{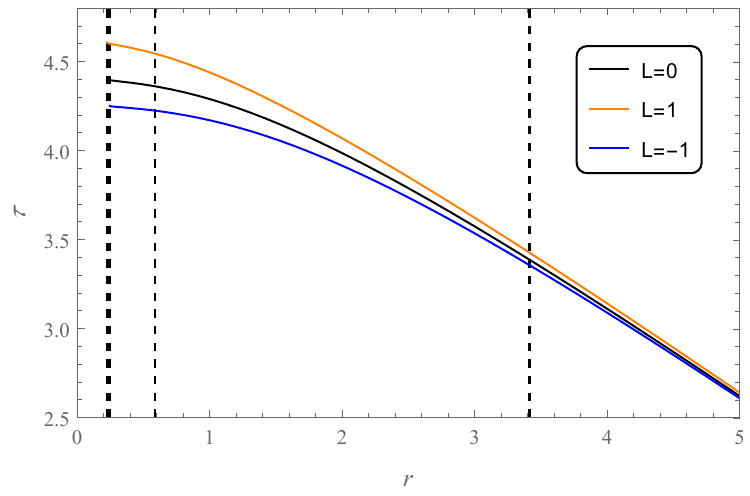}}}\quad
	\subfloat[\label{BH_time-vel}]{{\includegraphics[scale=0.55]{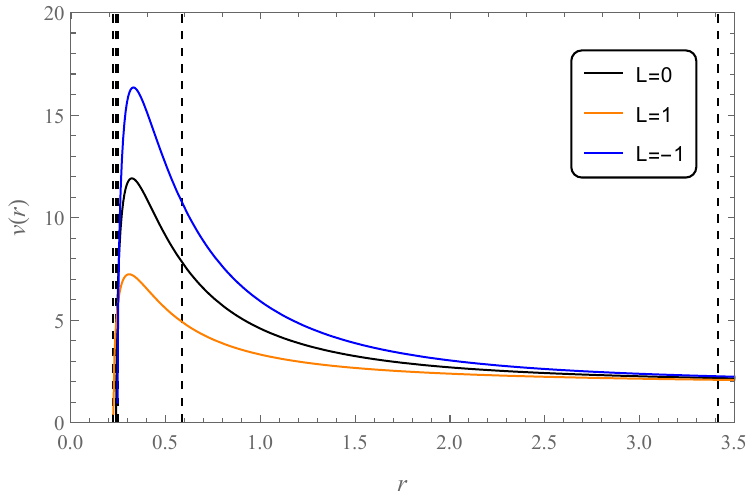}}}
	\caption{(a) Spacetime diagram of radially infalling particles in terms of proper time ($ \tau $) in KN black hole. For $ E=2,~\mu=1 $, and $ a=Q=1 $ and $ M=2 $, geodesics of zero and non-zero AM particles are shown along with the confinement points. (b) Velocity profiles of radially infalling zero and non-zero AM particles in KN black hole with $ E=2,~\mu=1 $, and $ a=Q=1 $ and $ M=2 $. The radial velocity with respect to proper time ($ \tau $) goes to zero as the particles hit the confinement. In both (a) and (b), the vertical dashed lines from left to right denote confinement points, inner and outer horizons respectively.}
	\label{BH_proper}
\end{figure}

Now, for the spacetime diagram of radial timelike geodesics, one may impose the black hole source parameter choices in Eq. \eqref{2.12} and \eqref{2.11}. For $ E=2,~\mu=1,~M=2 $, and $ a=Q=1 $, the plots are provided in the bottom panel of Fig. \ref{BH-fig} with separate frames for $ L=0,~L=1 $, and $ L=-1 $ motions.

Again, similar to the motion of photons, test particle trajectories show asymptotic nature leading to the infinite time consumption at the horizons. Although the particle trajectory confinements confirm the presence of positive AM test particles alone in CTC, the inbound motions of Fig. \ref{BH-fig} (bottom panel) are needed to be improvised in terms of proper time ($ \tau $). In this case, we directly use Eq. \eqref{2.8} for a timelike motion, by turning the numerator and denominator as
\begin{widetext}
	\begin{eqnarray}
		\frac{d\tau}{dr} = \frac{r^2}{\sqrt{(E^2-\mu)r^4+2Mr^3\mu-(-a^2E^2+L^2+a^2\mu+Q^2\mu)r^2+(-aE+L)^2(2Mr-Q^2)}}.
		\label{tau-eq}
	\end{eqnarray}
\end{widetext}
Employing numerical integration on this equation, one can readily obtain the radial geodesics without the inconsistency of coordinate time. For the same source parameters as Fig. \ref{BH-fig}, the plots for zero and non-zero AM particles are obtained in Fig. \ref{BHTG-proper}. Notice that the geodesics are well-behaved at the inner and outer horizons, and they now hit the confinement points with finite times. Although, as expected, their confinement points are the same as Fig. \ref{BH-fig} (bottom panel). The fact clarifies the limitation of coordinate time to describe the geodesic motion in black holes.

Further, to extend the characteristics of motion inside the black hole, we intend to carry out the velocity profile which has a particular point of interest for particles with different AM. In Fig. \ref{BH_time-vel}, the velocities of test particles for different AM are visualized along with the turning points where velocity ceases. At these points, in general, the velocity of each geodesic falls to zero, thus, the particles stop moving and confinements appear.

Speaking in terms of gravitational field, the infalling geodesics from the spatial infinity would approach towards the event horizon where attractive gravitational fields gradually increase. As they close near the event horizon, the increasingly higher gravitational field bends the spacetime curvature extremely strongly and the geodesics never escape. The coordinate singularity present in the horizons affects coordinate time as seen from Fig. \ref{BH-fig}. Let us consider the outer horizons in the figures. The infalling particle falls to the negative time infinity, although, as one gradually observes to the left, it rises from the negative infinity and again approaches the positive time infinity at the inner horizon. However, in terms of proper time, the inconsistency diminishes and particle velocity rises between the outer and inner horizons. It would feel a significantly higher gravitational field as it crosses the inner horizon and the field rises to maximum near the Cauchy surface where the velocity reaches the peak. Beyond that, the repulsive field of gravity starts to dominate, and the velocity falls rapidly before ultimately turning to zero at the confinement radius. It must be noted that particles are forbidden to interact with the central singular point both in black hole and naked singularity of Kerr-Newman spacetime. There must present a strong repulsive effect of gravity whose effect varies with the characteristics of particles, such as angular momentum.

\begin{figure*}[t]
	\centering
	\subfloat[\label{Ext-BH-fig1}]{{\includegraphics[scale=0.39]{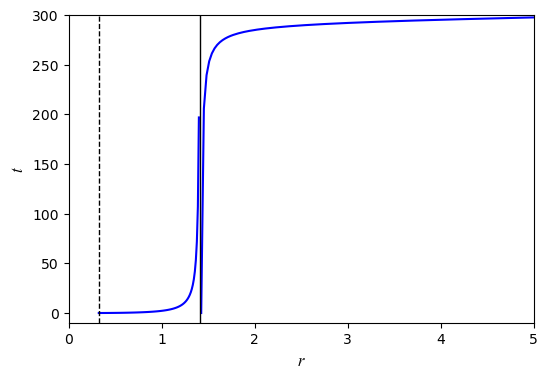}}}\qquad
	\subfloat[\label{Ext-BH-fig2}]{{\includegraphics[scale=0.39]{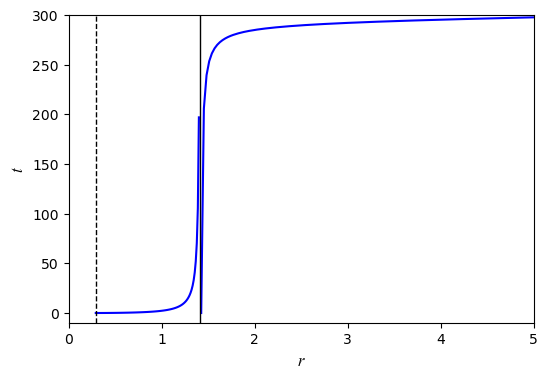}}}\qquad
	\subfloat[\label{Ext-BH-fig3}]{{\includegraphics[scale=0.39]{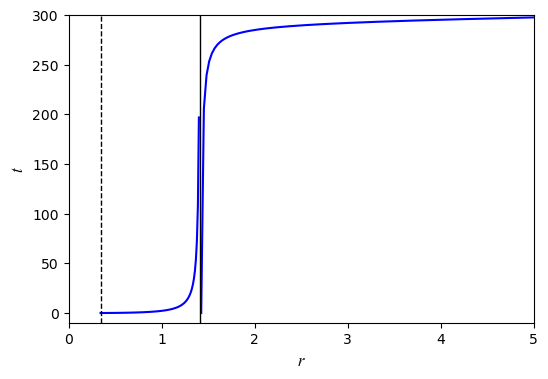}}}
	\caption{Spacetime diagram for radial timelike geodesics of KN extremal black hole with $ r_{ctc}=0.315 $ and $ r_{horizon}=M=1.414 $. The plots are exhibited for $ E=2,~\mu=1,~a=Q=1,~ M=\sqrt{2} $, and are separated with respect to angular momentum of particles as (a) $ L=0 $, (b) $ L=1 $, and (c) $ L=-1 $. The confinement points are shown in black dashed lines, and are located at $ r=0.324,~r=0.291,~ \text{and}~ r=0.343 $ respectively. The vertical solid lines define the event horizon.}
	\label{Ext-BH-fig}
\end{figure*}

Up to this point, we only considered two horizon non-extremal black holes, but what happens if we consider extremal black holes? One may rightfully check the position of confinement radius in the Kerr-Newman extremal black hole where the event horizon is described by $ r_H=M=\sqrt{a^2+Q^2} $, and understand the nature of particles in CTC. In this context, Fig. \ref{Ext-BH-fig} is presented to note the confinement radius for different AM timelike particles. The source parameters and particle energy are considered as $ M=\sqrt{2},~a=Q=1,~\mu=1 $ and $ E=2 $. The Cauchy surface appears at $ r=0.315 $, which is inside the event horizon at $ r_{horizon}=1.414 $. Referring to the figure, we can observe that similar to non-extremal black holes, infalling particles are always repelled before approaching the singularity and they are confined at the positions significantly far from $ r=0 $. Subsequently, only positive AM particles cross the radius of Cauchy horizon and only they are capable of traversing CTC.

\textbf{Confinement radius:} One of the promising aspects coming out of this discussion is the confinement of geodesic trajectories. The formula regarding this confinement radius is, in general, the point of interest, and it can be obtained directly from the real roots of the expression inside the square-root term of Eq. \eqref{2.8}. Although, finding the analytical solution is a difficult task to make, the roots are given by \footnote{Note that, this $ A $ and $ B $ is different from the $ A $ and $ B $ of effective potential, as given by Eq. \eqref{2.13}.}
\begin{eqnarray}
	r^{1}_{\pm} &\equiv& -\frac{B}{4A}-\frac{\sqrt{\delta+\chi}}{2} \pm \frac12 \sqrt{2\delta-\chi-\frac{\epsilon}{4\sqrt{\delta+\chi}}} ,
	\label{confine1} \\
	r^{2}_{\pm} &\equiv& -\frac{B}{4A}+\frac{\sqrt{\delta+\chi}}{2} \pm \frac12 \sqrt{2\delta-\chi+\frac{\epsilon}{4\sqrt{\delta+\chi}}} ;
	\label{confine2}
\end{eqnarray}
where
\begin{eqnarray}	
	\nonumber	\chi &=& \frac{2^{1/3}\alpha}{3A\left(\beta+\sqrt{-4\alpha^3+\beta^2}\right)^{1/3}} + \frac{\left(\beta+\sqrt{-4\alpha^3+\beta^2}\right)^{1/3}}{3 \times 2^{1/3}A} ,\\
	\nonumber	\delta &=& \frac{B^2}{4A^2}+\frac{2F}{3A} ,\\
	\nonumber	\epsilon &=& -\frac{B^3}{A^3}-\frac{4BF}{A^2}-\frac{8G}{A} ,\\
	\nonumber	\alpha &=& F^2-3BG-12AH ,\\
	\nonumber	\beta &=& -2F^3 +9BFG+27AG^2-27B^2H-72AFH ,\\
	\nonumber	A &=& E^2-\mu ,\\
	\nonumber	B &=& 2 M \mu ,\\
	\nonumber	F &=& -a^2 E^2 +L^2 +(a^2+Q^2) \mu ,\\
	\nonumber	G &=& 2M (-aE+L)^2 ,\\
	\nonumber	H &=& Q^2 (-aE+L)^2 .
\end{eqnarray}
Now, depending on the choices of parameters, Eq. \eqref{confine1} and \eqref{confine2} always provide two complex roots, one positive definite and one negative definite real root. Here, only the positive real root is physically convenient, and is explicitly used throughout the study to accurately determine the confinement radius of geodesics in both the black hole and naked singularity. However, mostly $ r^1_+ $ provides the positive definite confinement radius.

\begin{figure*}[t]
	\centering
	\subfloat[\label{NS_r-L}]{{\includegraphics[scale=0.55]{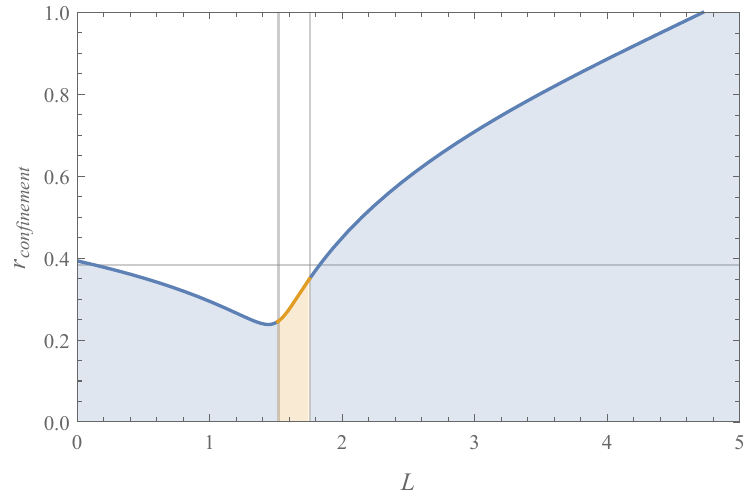}}}\quad
	\subfloat[\label{BH_r-L}]{{\includegraphics[scale=0.55]{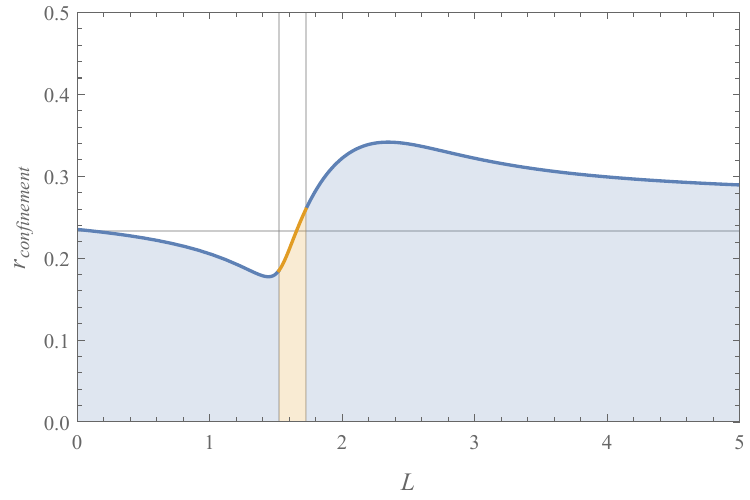}}}
	\caption{(a) Variation of confinement radius with angular momentum of test particles for $ \mu=1,~M=Q=1,~a=0.5 $ and $ E=3.2 $. The confinement radius $ r_{confinement} $ is minimum i.e. $0.23809$, at $ L=1.4437 $, creating a void surrounding the singularity. (b) Equivalent plot for black hole interior geometry with  $ \mu=1,~M=2,~Q=1,~a=0.5 $ and $ E=3.2 $, where $ r_{confinement} $ is minimum i.e. $0.17735$, at $ L=1.4437 $. The blue and orange lines in both the plots respectively denote $ r^1_+ $ and $ r^2_+ $ and the horizontal gridline represents the CTC radius.}
	\label{r-L_plot}
\end{figure*}

Now, let us have a closer look at the minimum confinement radius (or maximum radial reach of the particles towards the singularity). In this context, it is necessary to report the critical point $ aE=L $ for which $ r^1_+ $ terminates for photons. However, we may visualize the dependence of confinement radius with the angular momentum of test particles by putting other parameters fixed. The result for naked singularity is shown in Fig. \ref{NS_r-L}. It confirms the empty region around the singularity where the minimum confinement radius is registered at $ L=1.4437 $, and the value of $ r^1_+ $ is complex around both directions of $ L=aE=1.6 $ from $ L=1.52 $ to $ L=1.755 $. Interestingly, this gap from $ L=1.52 $ to $ L=1.755 $ is the region where $ r^2_+ $ changes value from complex to real definite. Thus, we have figured out the missing ingredient and plotted them in \ref{NS_r-L} using $ r^2_+ $. Additionally, it is observed that for $ L>aE $, the radius gradually increases and goes larger than the CTC radius after a certain value representing the unavailability of particles in the CTC with those values of $ L $.

Fig. \ref{BH_r-L} exhibits the corresponding plot for black holes. Here, $ r^1_+ $ attains complex values within $ L=1.522 $ to $ L=1.73 $, and the confinement radius is minimum i.e. $ 0.17735 $ at $ L=1.4437 $. Similar to the naked singularity, the figure is completed by plotting $ r^2_+ $ within $ L=1.522 $ to $ L=1.73 $. Furthermore, one may consider the value (of angular momentum) $ L=1.4437 $ to be a critical one for $ a=0.5 $, and $ E=3.2 $ for both the naked singularity and black hole. However, a key point coming out from this investigation is that positive angular momentum is not always the sufficient condition to let a particle traverse the CTC. If we carefully notice, we may find that for a significantly small positive value of angular momentum (that means $ L<<aE $) in a timelike particle, it cannot pass through the Cauchy surface. On the other hand, the result is the same for any timelike particle with $ L>aE $ in a black hole interior (possible for only a certain value in naked singularity). 

\section{Motion of charged particles}\label{charged}
Since, the dynamics of neutral test particles have been extensively investigated, we are now shifting our attention into the notion of charged particles in a Kerr-Newman background. The Lagrangian density provided by the test particle of mass $ \mu $ and charge $ q $ moving around the background described by the line element of Eq. \eqref{2.2} is given by
\begin{equation}
	\mathcal{L}=\frac12 g_{\alpha \beta} \dot{x}^{\alpha} \dot{x}^{\beta} + \epsilon A_{\alpha} x^{\alpha},
	\label{ch01}
\end{equation}
where $ A_{\alpha} $ denotes the components of electromagnetic 4-potential having $ A=\frac{Q}{r} dt $, and $ F=dA=-\frac{Q}{r^2} dt \wedge dr $. Here, the overdot represents differentiation with respect to proper time, and the specific charge is $ \epsilon=q/ \mu $. One can now readily compute the equations of motion according to the Euler-Lagrange equation, such that
\begin{equation}
	\dot{x}^\alpha \nabla_\alpha \dot{x}^\beta = \epsilon F^\beta_\gamma \dot{x}^\gamma.
	\label{ch02}
\end{equation}

Then, after few lines of calculations, we derive the geodesic equations in terms of conserved quantities ($ L $ and $ E $), and are given by
\begin{widetext}
	\begin{eqnarray}
		\dot{t} &=& - \frac{aL(2Mr-Q^2)}{r^2(r^2-2Mr+a^2+Q^2)} + \frac{r^4+a^2(r^2+2Mr-Q^2)}{r^2(r^2-2Mr+a^2+Q^2)} \left( E+\frac{\epsilon Q}{r} \right),
		\label{ch03} \\
		\dot{\phi} &=& \frac{L(r^2-2Mr+Q^2)}{r^2(r^2-2Mr+a^2+Q^2)} + \frac{a(2Mr-Q^2)}{r^2(r^2-2Mr+a^2+Q^2)} \left( E+\frac{\epsilon Q}{r} \right),
		\label{ch04} \\
		\nonumber
		\dot{r}^2 &=& \frac{1}{r^6 \mu^2} \Big[(Er \mu+qQ)^2(r^4+a^2(r^2+2Mr-Q^2)) 
		+2aL \mu r(Er \mu+qQ)(Q^2-2Mr) 
		\\
		&&+ r^3 \mu(r \epsilon \mu-2qQ)(r^2-2Mr+a^2+Q^2) - L^2r^2 \mu^2(r^2-2Mr+Q^2) \Big].
		\label{ch05}
	\end{eqnarray}
	
	Here, we are skipping the derivations of effective potential which can be obtained by a sufficiently easy method, or by repeating the steps as discussed in section \ref{geodesics}. So, we move our attention directly to the expressions required to obtain the spacetime diagram. Following the division of Eq. \eqref{ch03} by \eqref{ch05}, the expression takes the form:
	\begin{eqnarray}
		\nonumber
		\frac{\dot{t}}{\dot{r}}= \left[ \frac{(r^4+a^2(r^2+2Mr-Q^2))(Er+\epsilon Q)-aLr(2Mr-Q^2)}{(r^2-2Mr+a^2+Q^2)}\right]\mu \bigg/ \biggl[(Er \mu+qQ)^2(r^4+a^2(r^2+2Mr-Q^2)) 
		\\
		+2aL \mu r(Er \mu+qQ)(Q^2-2Mr) 
		+ r^3 \mu(r \epsilon \mu-2qQ)(r^2-2Mr+a^2+Q^2) - L^2r^2 \mu^2(r^2-2Mr+Q^2) \biggr]^{1/2}.
		\label{ch06}
	\end{eqnarray}
\end{widetext}

\begin{figure*}[t]
	\centering
	\subfloat[\label{charged-fig1}]{{\includegraphics[scale=0.4]{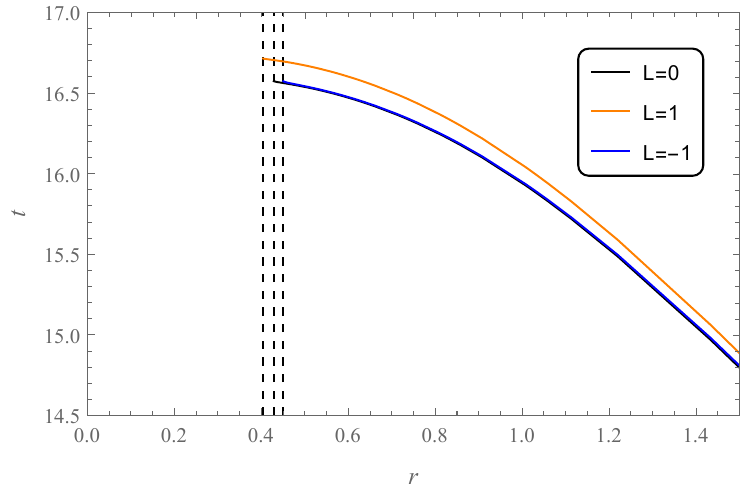}}}\qquad
	\subfloat[\label{charged-fig2}]{{\includegraphics[scale=0.4]{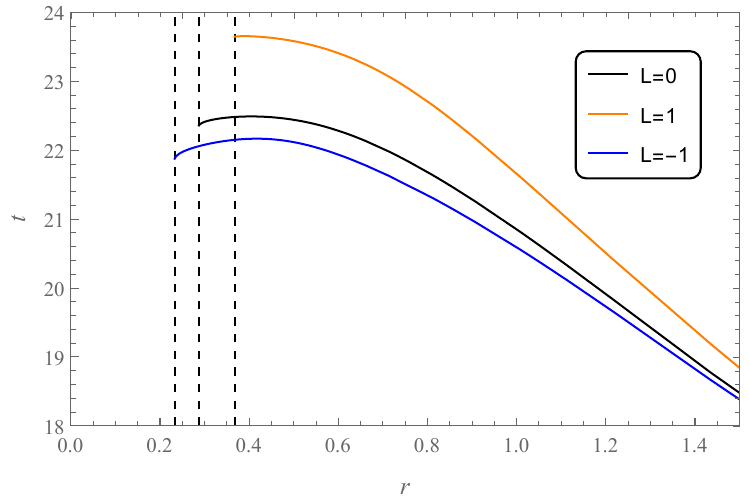}}}\\
	\subfloat[\label{charged-fig3}]{{\includegraphics[scale=0.39]{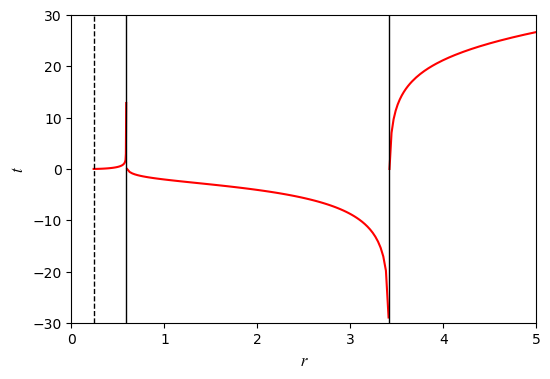}}}\qquad
	\subfloat[\label{charged-fig4}]{{\includegraphics[scale=0.39]{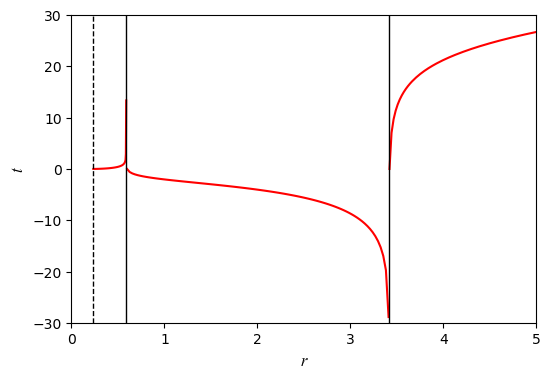}}}\qquad
	\subfloat[\label{charged-fig5}]{{\includegraphics[scale=0.39]{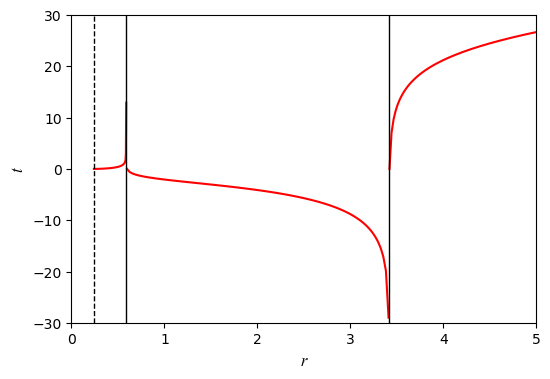}}}\\
	\subfloat[\label{charged-fig6}]{{\includegraphics[scale=0.39]{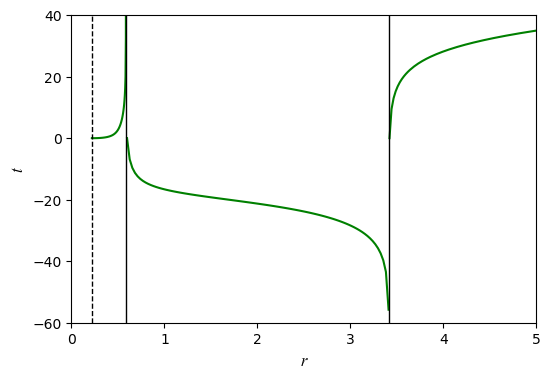}}}\qquad
	\subfloat[\label{charged-fig7}]{{\includegraphics[scale=0.39]{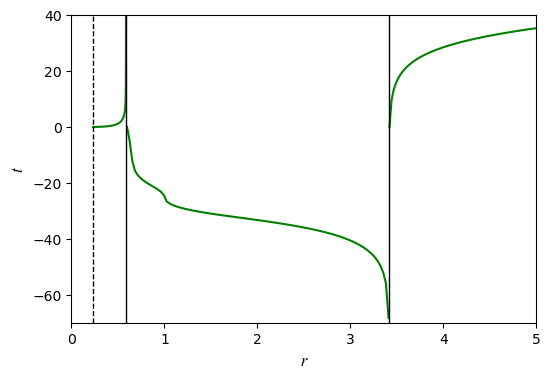}}}\qquad
	\subfloat[\label{charged-fig8}]{{\includegraphics[scale=0.39]{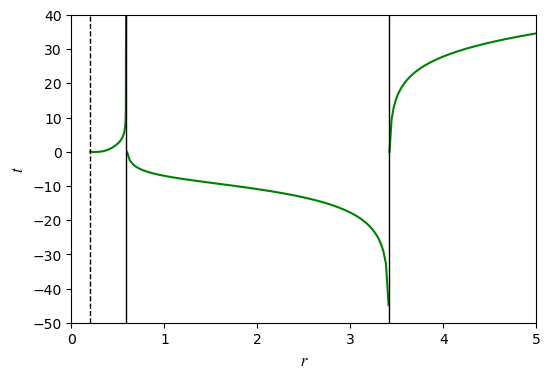}}}
	\caption{TOP PANEL: The spacetime diagram of charged test particles in naked singularity. (a) The trajectories of positively charged particles ($ q=+1 $) with zero and non-zero AM are plotted with $ M=a=Q=1,~\mu=\epsilon=1 $ and $ E=2 $. The $ L=0,~L=1 $ and $ L=-1 $ plots are respectively confined at $ r=0.428,~r=0.4039 $ and $ r=0.450 $. (b) For $ q=-1 $, the confinements are $ r=0.287,~r=0.367 $ and $ r=233 $.
		\newline MIDDLE PANEL: Spacetime diagram of positively charged particles ($ q=+1 $) with $ M=2,~a=Q=1,~\mu=\epsilon=1 $ and $ E=2 $ in non-extremal black hole. The (c) $ L=0 $, (d) $ L=1 $, (e) $ L=-1 $ motions are respectively confined at $ r=0.239,~r=0.2351,~ r=0.242 $ and are shown in vertical dashed lines.
		\newline BOTTOM PANEL: Spacetime diagram of negatively charged particles ($ q=-1 $) with $ M=2,~a=Q=1,~\mu=\epsilon=1 $ and $ E=2 $ in non-extremal black hole. The (f) $ L=0 $, (g) $ L=1 $, (h) $ L=-1 $ motions are respectively confined at $ r=0.22,~r=0.233,~ r=0.201 $ and are shown in vertical dashed lines. The CTC and the event horizons (vertical solid lines) are located at $ r_{ctc}=0.235,~ r_{in}=0.586 $ and $ r_{out}=3.414 $.}
	\label{charged-fig}
\end{figure*}

The associated numerical integration of this equation is performed to visualize the spacetime diagrams as shown in Fig. \ref{charged-fig}. For the choices of source parameters, we have picked the same values from sections \ref{naked-sing} and \ref{black-hole} to examine the changes that a charged particle makes. It is not surprising to note that for positive charges of the test particles i.e. $ q=+1 $ (with source charge $ Q=+1 $), the phenomenon is not different from the neutral test particles. The gravitational force for neutral particles in both the naked singularity and black hole, sourced from the inner region of the geometry and an infalling particle from spatial infinity would feel it increasing higher and higher as the particle gradually moves towards the singularity. The presence of positive AM for which the particle co-rotate with the spacetime, provides an additional push so that it survives the repulsive gravitation more in the inner region and thus, they move more towards the singularity. The scenario is much more compelling in the presence of a positive charge in the particles. This is a straightforward example where the gravitational interaction through the angular rotation of particles and the electromagnetic interaction acts together. For the presence of Coulomb interaction, the particles feel an additional repulsion, and the effect of gravitation embedded in the angular rotation although survived, the confinement radii move outwards except for the counter-rotating (i.e. negative AM) particles. For negative AM particle motion, surprisingly the gravitational repulsion and the electromagnetic repulsion acts oppositely and the confinement radii move inwards. Here, identical charge although repels the counter-rotation generates an opposite inward pull that disorients the gravitational repulsion, and thus, the confinements move inward. However, the particles with positive AM still manage to cross the Cauchy radius and traverse the CTC. For naked singularity, $ L=1 $ particles now reach up to $ r=0.4039 $ radial position where CTC appears inside $ r=0.4047 $. The same particles in black holes reach up to $ r=0.2351 $ which is inside the CTC that is possible at $ r<0.2354 $.

For negatively charged particles ($ q=-1 $) in the background of a positively charged source ($ Q=+1 $), the behavior is completely opposite as compared to neutral particles. Here, particles are heavily attracted towards the inner region and the negative AM particles are those who are attracted the most. From the physics of Coulomb attraction and repulsion, the identical charged particles that were repelled more will be attracted more when the charges are opposite. Thus, for both the naked singularity and black hole, all the confinement radii are significantly pulled inward, and unlike the neutral scenario, the confinement of negative AM particles lie in the innermost region, then AM-less particles, and then positive AM particles in the outermost position. This phenomenon indicates that the Coulomb interaction dominates, and outweighs the gravitational interaction that was dominating one in the dynamics of neutral particles. Again, a key point is that, all the particles are now capable of traversing the closed timelike curves in the naked singularity and black hole. Comparing the position of CTC (which appears inside $ r=0.4047 $ and $ r=0.2354 $ radius in NS and BH respectively) with the confinement radius given in the bottom panel of Fig. \ref{charged-fig}, one may verify the given results.

\begin{figure*}[t]
	\centering
	\subfloat[\label{Ext-BH_charge-fig1}]{{\includegraphics[scale=0.39]{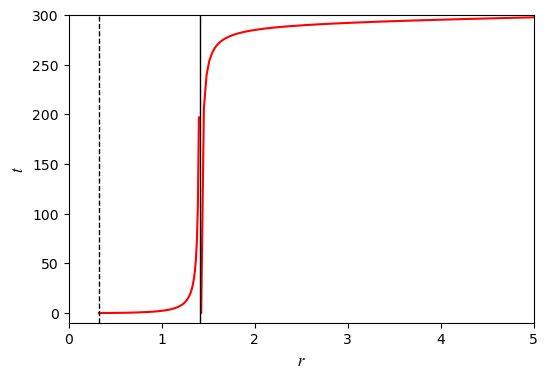}}}\qquad
	\subfloat[\label{Ext-BH_charge-fig2}]{{\includegraphics[scale=0.39]{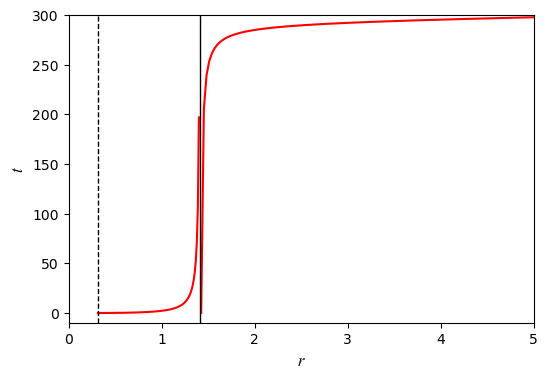}}}\qquad
	\subfloat[\label{Ext-BH_charge-fig3}]{{\includegraphics[scale=0.39]{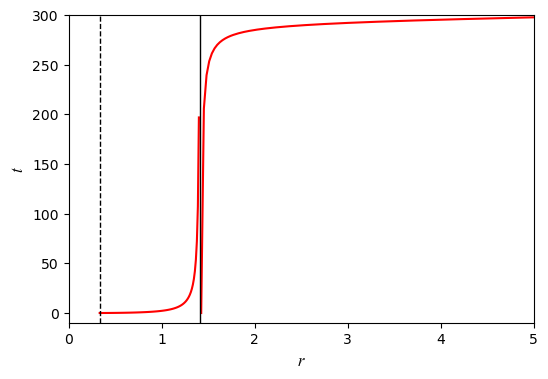}}}\\
	\subfloat[\label{Ext-BH_charge-fig4}]{{\includegraphics[scale=0.39]{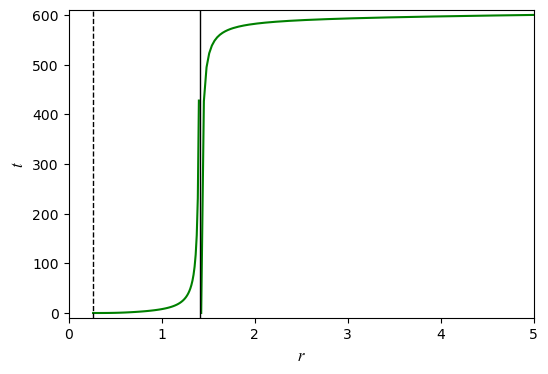}}}\qquad
	\subfloat[\label{Ext-BH_charge-fig5}]{{\includegraphics[scale=0.39]{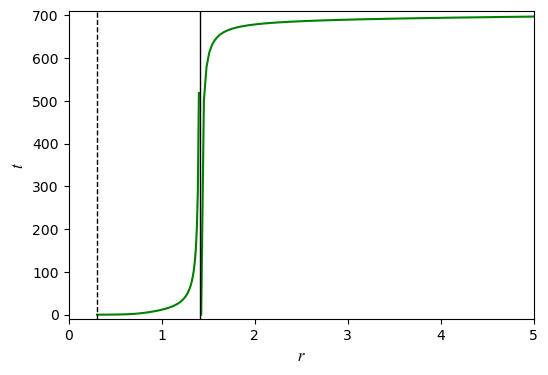}}}\qquad
	\subfloat[\label{Ext-BH_charge-fig6}]{{\includegraphics[scale=0.39]{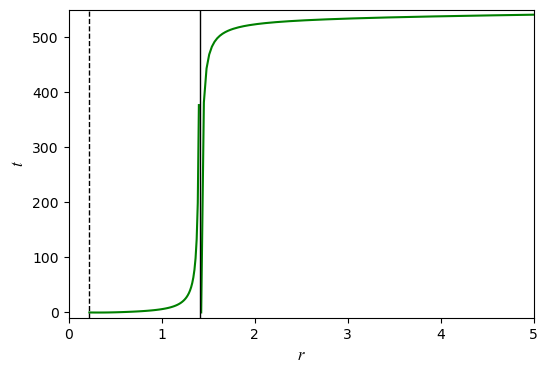}}}
	\caption{TOP PANEL: Spacetime diagram of positively charged particles ($ q=+1 $) with $ M=\sqrt{2},~a=Q=1,~\mu=\epsilon=1 $ and $ E=2 $ in KN extremal black hole. The (a) $ L=0 $, (b) $ L=1 $, (c) $ L=-1 $ motions are respectively confined at $ r=0.325,~r=0.3144,~ r=0.333 $ and are shown in vertical dashed lines.
		\newline BOTTOM PANEL: Spacetime diagram of negatively charged particles ($ q=-1 $) with $ M=\sqrt{2},~a=Q=1,~\mu=\epsilon=1 $ and $ E=2 $ in KN extremal black hole. The (d) $ L=0 $, (e) $ L=1 $, (f) $ L=-1 $ motions are respectively confined at $ r=0.262,~r=0.305,~ r=0.222 $ and are shown in vertical dashed lines. The CTC and the event horizon (vertical solid line) are located at $ r_{ctc}=0.315 $ and $ r_{horizon}=1.414 $ respectively.}
	\label{Ext-BH_charge-fig}
\end{figure*}

The confinement of geodesics in Kerr-Newman extremal black holes has also been investigated in the presence of a positive ($ q=+1 $) and a negative ($ q=-1 $) charge within the particles. For exactly same source parameter choices as the neutral particle case i.e. $ M=\sqrt{2} $ and $ a=Q=1 $, we obtained the confinement radius of the charged particles as shown in Fig. \ref{Ext-BH_charge-fig}. Here, it is readily observed that the discussions on the origin of confinements in non-extremal black holes also hold in extremal KN black holes. For identical charged particles, the confinement radii move outwards due to Coulomb repulsion (except for negative AM geodesics), and only positive AM particles are allowed in the CTC. However, for opposite charged particles, all the confinement boundaries are strongly pulled inside by Coulomb attraction, and each particle is allowed to be present in the neighborhood of CTC.

Finally, one can notice that the charge conservation law is valid in the context of charged particle motion. If we reverse the sign of charge in the source (i.e. if we consider $ Q=-1 $), then all the characteristics are reversed. Namely, the nature of the plots and the confinement points of $ Q=-1,~q=+1 $ will be same as $ Q=+1~q=-1 $; and $ Q=-1,~q=-1 $ will be same as $ Q=+1~q=+1 $.

Further, for the formation of CTC, the azimuthal coordinate $ \phi $ should be timelike in nature keeping all other coordinates (i.e. $ t,~r,~\theta $) constant. Therefore, from Eq. \eqref{2.2}, we have
\begin{equation}
	ds^2=\frac{r^4+a^2(r^2+2Mr)-a^2Q^2}{r^2}d\phi^2.
\end{equation}
Hence, for $\phi$ to be a timelike coordinate, the source charge $ Q $ has a dominating role and relation \eqref{2.5} must be satisfied. The region of formation of CTC is thus given by $ r<r^-_+ $. So, from the Lagrangian formulation $ \mathcal{L}=\frac12 g_{\alpha \beta} \dot{x}^{\alpha} \dot{x}^{\beta} $, we have $ 2\mathcal{L}=\bar{A} \dot{\phi}^2 $ in the region of CTC, where $ \bar{A}=\frac{r^4+a^2(r^2+2Mr)-a^2Q^2}{r^2} $. Therefore, the corresponding angular momentum ($ L_c $) of a timelike particle will be
\begin{equation}
	L_c= \bar{A} \dot{\phi},
\end{equation}
and it is conserved. As $ \bar{A} $ is negative in the CTC region (to make the angular coordinate $ \phi $ to be timelike), the angular velocity $ \dot{\phi} $ of any test particle depends on the angular momentum of the test particles as follows:

$\bullet$ The test particle will have counter-rotating angular velocity when angular momentum is positive $ L_c>0 $, and hence it favors the formation of CTC, since CTC counter-rotates with the source spin i.e., the time orientation of the CTC is opposite to the direction in which the singularity rotates. For detailed discussion refer to \cite{Duan:2021pci}.

$\bullet$ The test particle will have co-rotating angular velocity for negative angular momentum particles $ L_c<0 $, and it is not favorable for the formation of CTC.

Note that this analysis is independent of the charge of test particles. Moreover, the above analysis does not imply that timelike geodesics or timelike curves within the region $ r<r^-_+ $ does not necessary to have violation of causality (for details, see references \cite{Clement:2015cxa,Clement:2015aka,Clement:2022pjr}).

\section{Conclusions}\label{discussions}

In this study, we performed a detailed analysis of the existence of closed timelike curves and the characteristics of test particles along their orbits on the equatorial slice of KN spacetime.
In this context, first we investigated the position of CTC in the orbital region of the geometry, and found two real roots where only one of them is positive and physically realizable. The graphical representation of the inner horizon, outer horizon, and the region of CTC showed that the Cauchy surface is always located inside the inner horizon, and the central singularity at $ r=0 $ is covered by CTCs, both in the black hole and naked singularity.

The geodesic equations give rise to the existence of spacetime dragging effect in the zero AM test particles. However, together with the positive and negative AM particles which respectively co-rotate and counter-rotate with the spacetime spin, these three infalling geodesics ultimately terminate at the confinement point that restricts them to approach the central singular point at $ r=0 $. So, there exists an empty region surrounding the singularity where particle movements are completely forbidden. Referring to \cite{Pugliese:2013zma}, Pugliese \textit{et al.} analyzed that, in terms of circular motion, the radius of the forbidden area ($ r_*=Q^2/M $) which covers the singularity, situated inside the outer horizon. They also argued that although it does not depend on the source parameters, it is completely a property of the gravitational field generated by the electric charge. In RN, this boundary corresponds to the limiting radius of zero AM particles \cite{Pugliese:2010ps}. However, in the present study, the visualization is getting more transparent. In both the black hole and naked singularity, although the presence of an empty region surrounding the singular point makes it inaccessible for test particles, the radius of the region certainly depends both on the source parameters and the characteristics of test particles (such as energy and total angular momentum). We obtain the explicit expression for this radius as given by Eq. \eqref{confine1} and \eqref{confine2}, which also represent the limiting radius of the positive AM particles for neutral and identical charged particle (with the source charge) motions; and for the opposite charged particles, this is the limiting radius of negative AM particles. Regardless of the confinement of different AM test particles, the empty region lies way inside the inner horizon for black holes. However, for the oppositely charged particles, the confinements move slightly closer to the central singularity, as compared to the neutral ones.

As discussed, for neutral and identical charged particle motions, an observer sitting on the singularity first sees the positive AM particle region where CTCs exist. Particles can revolve in these curves up to the zero AM confinement radius where only zero and positive AM particles exist. Finally, after the negative AM confinement, all three types of particles are free to exist. For opposite charged particle motion, the observer respectively sees the confinement of negative AM, zero AM and positive AM particles. Due to the Coulomb attraction, however, they all are able to traverse the CTC altogether.

Apart from the spacetime diagram, the motion of particles both in the black hole and naked singularity are also investigated in terms of velocity and/or the effective potential of the timelike geodesics. The infalling geodesics feel an increasingly higher gravitational field as it gradually approaches the singularity. Thus, the velocity rises consistently up to a certain radius (which is close to the singularity) where repulsive gravity effects start to dominate and the velocity falls to zero preventing the particles to further approach the singularity. The increasingly higher order of repulsive gravity in the vicinity of the central singularity hides it from outside and prevents us from studying the singular point with particle motions.

The motion of test particles in black hole always has a particular point of interest. As established from the geodesic structures, the coordinate time has certain drawbacks in defining the infalling geodesics at the coordinate singularity; still, it indicates the presence and maximum radial reach of the test particles inside the outer and inner horizons. Further, we considered the infalling trajectories in terms of proper time in the geometry, and establish the results along with the characteristics of particles in CTCs inside the inner horizon.

For the motion of neutral particles, the zero AM photon confinement radius is the same as the boundary of CTC, e.g. referring to Fig. \ref{NS_spacetime-fig}, the geodesics of zero AM particles are only possible at $ r>0.405 $, where CTCs exist at $ r<0.405 $ radius. Further, it is only the positive AM particles that can reach $ r<0.322 $, i.e. inside the region of CTC. So, it is evident that only the positive AM particles can exist in the neighborhood of CTC in the naked singularity. The result is however also true for KN black holes as discussed in section \ref{black-hole} where one can refer to Fig. \ref{BH-fig}, and \ref{BH_proper} for such an example. Thus, CTCs only permit particles with positive AM, thereby restricting particles with zero and negative AM.

The motion of charged particles in section \ref{charged} partially connects sections \ref{naked-sing} and \ref{black-hole}, especially concerning the types of particles capable of traversing CTCs. When an infalling charged particle follows its trajectory in the Kerr-Newman (KN) geometry, it doesn't yield any surprising differences than neutral ones. For identical charged particles (with the same source charge), CTC still permits only the positive AM particles in the vicinity, albeit with the confinement radius primarily shifting outward. However, for oppositely charged particles, the strong Coulomb attraction outweighs the gravitational interaction, causing all three types of particles to be pulled inside the radius of Cauchy surface. As a result, they all become capable of traversing the CTC.

Thus, in this present work, not only the characteristics of particles in CTC is studied, but also it shed light on the motion of particles in the equatorial orbit of KN black hole and naked singularity. Though, in the last few years, much efforts were made to understand the formation, structure and existence of naked singularity \cite{Patil:2010nt,Joshi:2011rlc,Patil:2011uf,Cohen:1979zzb,DiCriscienzo:2010gh,liang1974,Dotti:2008ta,Casadio:2003iv,Toth:2012vvy,Virbhadra:2007kw,Virbhadra:2002ju}, still, these studies are confined to investigate the motion of particles in terms of differentiating the accretion disk formations of black hole and naked singularity over spinning spacetime \cite{Babichev:2008dy,stuchlik1999,stuchlk1996,stuchlik1998,kovr2007,esteban1990,jaroszynski1980}. In the present work, the motion of test particles in KN spacetime may contribute some insight to the study and discussions of naked singularity and the interior structure of black hole as well. Finally, one may conclude that the mere existence of CTCs may not necessarily signal an observed violation of causality.

\section*{Acknowledgement}
The authors thank the anonymous referee whose comments and suggestions improved the quality and visibility of the paper. The authors are also thankful to the Inter University Centre for Astronomy and Astrophysics (IUCAA), Pune (India) for their warm hospitality, as a part of this work was carried out there during a visit. S.C. thanks FIST program of DST, Department of Mathematics, JU (SR/FST/MS-II/2021/101(C)).

\end{document}